\documentclass[aps,prd,twocolumn,superscriptaddress]{revtex4}
\usepackage[english]{babel}
\usepackage[latin1]{inputenc}
\usepackage{graphicx,color}
\usepackage{latexsym}
\usepackage{amsmath}
\usepackage{amsfonts}
\usepackage{amssymb}
\usepackage[latin1]{inputenc}

\def\bb{\bibitem}

\def\bb{\bibitem}

\newcommand{\be}{\begin{equation}}
\newcommand{\ee}{\end{equation}}
\newcommand{\bea}{\begin{eqnarray}}
\newcommand{\eea}{\end{eqnarray}}

\begin{document}

\title{The geometric phases in a background with Lorentz-symmetry violation}

\author{L. R. Ribeiro}
\affiliation{{ Departamento de
F\'{\i}sica, Universidade Federal da Para\'\i ba, Caixa Postal 5008, 58051-970,
Jo\~ao Pessoa, PB, Brasil}}

\author{E. Passos}
\email{passos,lrr,furtado,jroberto@fisica.ufpb.br}
\affiliation{{ Departamento de
F\'{\i}sica, Universidade Federal da Para\'\i ba, Caixa Postal 5008, 58051-970,
Jo\~ao Pessoa, PB, Brasil}}

\author{C. Furtado}
\affiliation{{ Departamento de
F\'{\i}sica, Universidade Federal da Para\'\i ba, Caixa Postal 5008, 58051-970,
Jo\~ao Pessoa, PB, Brasil}}

\author{J. R. Nascimento}
\affiliation{{ Departamento de
F\'{\i}sica, Universidade Federal da Para\'\i ba, Caixa Postal 5008, 58051-970,
Jo\~ao Pessoa, PB, Brasil}}
\affiliation{Instituto de F\'\i sica, Universidade de S\~ao Paulo\\
Caixa Postal 66318, 05315-970, S\~ao Paulo, SP, Brazil}

\begin{abstract}
We analyze the nonrelativistic quantum dynamics of a single neutral spin half particle, with non-zero magnetic and electric dipole moments, moving in an external electromagnetic field in presence of Lorentz symmetry violation background. Also, we study the geometric phase for this model taking into account the parameter that breaks the Lorentz symmetry. We investigate the He-McKellar-Wilkens effect in this context and we verify that Lorentz violation parameter contribute to increase the electric dipole moment of the particle.
\end{abstract}

\maketitle

Studying the quantum dynamics of charged and neutral particles, in presence of electric and magnetic fields, we may realize several topological and geometrical effects. Aharonov and Bohm demonstrated that a charged particle circulating around a long and thin solenoid acquires a quantum topological phase in its wave function \cite{aha}. In the Refs.\cite{prl:cham,pes1}, this effect was observed experimentally. Aharonov and Casher showed that a neutral particle that possesses a non null magnetic dipole moment presents a quantum geometrical phase in its wave function while circulating around and parallel to a charged wire \cite{cas}. This effect was observed in a neutron interferometer \cite{cim} and in a neutral atomic Ramsey interferometer \cite{san}. He and McKellar \cite{mac}, and Wilkenes \cite{wil} independently, have obtained the quantum phase acquired by the wave function of a neutral particles, which have a non zero electric dipole moment, while it is circulating around and parallel to a line of magnetic monopole.  Anandan have investigated the general quantum phase acquired by a neutral particles that possess magnetic and electric dipole  in the presence of a external electromagnetic fields. Recently, one of us has demonstrated  that the phase of Anandam is a geometric quantum phase \cite{furtpra}. The proposed to
experimental scheme are given in the Refs.\cite{wei,franson}. Another point that has been discussed in the Ref.\cite{franson} is the unified description of all three effects. Beside, new effect called the dual Aharonov-Bohm effect is presented in the Refs.\cite{franson,furtadodual}. 

The possibility of Lorentz symmetry breaking was suggested in the Refs. \cite{0,1,2,3}. One of the implications of the Lorentz symmetry breaking is the possibility
of arising of new classes of couplings in the Lagrangian which involve constant vectors or tensors.
In the sense to detect the Lorentz-symmetry violation, many problems have been investigated in the last years. For instance, \v{C}erenkov-type mechanism called``vacuum \v{C}erenkov radiation'' to test the Lorentz symmetry \cite{ptl}, changing of gravitational redshifts for differently polarized Maxwell-Chern-Simons photons \cite{kkl}, evidence for the Lorentz and CPT  symmetries violation from the measurement of CMB polarization \cite{bjx}, \cite{hle} and magnetic monopoles inducing electric current \cite{brz},  aspect nonrelativistic for Lorentz-symmetry violation \cite{Kost1}, study non-minimal coupling by topological implications \cite{H1}, hydrogen and anti-hydrogen spectroscopy \cite{vl1,vl2}, observations in spin polarized torsion pendulum behavior \cite{vl3,vl4}, study sensitive to small Lorentz-symmetry violating effects from particle interference \cite{McKeller}.

In this paper we are interested in study the geometric phase to relativistic quantum dynamics of a single neutral spin half particle, with non-zero magnetic and electric dipole moments, moving in a external electromagnetic field in presence of approach to Lorentz-symmetry violation term. This approach, consists of adopting the Chern-Simons-like term ${\cal{L}_{CS}}=\frac{1}{2}\upsilon_{\alpha}\epsilon^{\alpha\beta\mu\nu}F_{\mu\nu}A_{\beta}$, with $\upsilon_{\alpha}$ being
a constant quadrivector characterizing the preferred direction of the space-time, to the photon sector \cite{1}. The nonrelativistic Hamiltonian associate to this system is obtained using the Foldy-Wouthuysen (FW) transformation for Dirac spinor. Also, the Aharonov-Casher and He-McKellar-Wilkens phases are analyzed in presence of a Lorentz symmetry violation parameter.

The system that we are interested is described by the following motion equation (we used $\hbar=c=1$):
\bea
\Bigl(i\gamma_\mu\partial^\mu+\frac12\mu\sigma_{\alpha\beta}F^{\alpha\beta}-\frac{i}{2}d\sigma_{\alpha\beta}\gamma_5F^{\alpha\beta}\nonumber\\+ ig \upsilon_{\alpha} {^*}F^{\alpha\beta}\hat{\beta}\gamma_{\beta}-m\Bigl)\psi=0\;,\label{eq:01}
\eea
where $\mu$ is the magnetic dipole moment, $d$ is the electric dipole moment, $g$ the coupling constant and $\upsilon_{\alpha}$ is the constant four vector which selects a fixed direction in space-time, and explicitly violates Lorentz and CPT symmetries.
We use the following convention by introduction an antisymmetric tensor of rank 2, the field-strength tensor $F^{\mu\nu}$ defined as
\begin{eqnarray*}
\qquad F^{\mu\nu}=\{\vec{E},\vec{B}\},\;\; F^{\mu\nu}=-F^{\nu\mu},\;\;\;F^{0i}=-E^i\\
F^{i0}=E^i,\;\; F^{ij}=-\epsilon_{ijk}B^k\;
\end{eqnarray*}
\begin{equation}
F^{\mu\nu}=\left[\begin{array}{cccc}
0 & -E^1 & -E^2 & -E^3\\
E^1 & 0 & -B^3 & B^2\\
E^2 & B^3 & 0 & -B^1\\
E^3 & -B^2 & B^1 & 0
\end{array}\right]\;
\end{equation}
The dual field-strength tensor ${^*}F^{\mu\nu}$ is obtained by contracting $F^{\mu\nu}$ with the completely antisymmetric unit tensor (the Levi-Civita tensor) $\epsilon^{\mu\nu\rho\sigma}$:
\be
{^*}F^{\mu\nu}=\frac{1}{2}\epsilon^{\mu\nu\rho\sigma}F_{\rho\sigma}=F^{\mu\nu}(\vec{E}\rightarrow \vec{B},\;\vec{B}\rightarrow -\vec{E}).
\ee
We have also,
\bea
&&\sigma_{0i}=\frac{i}{2}[\gamma_{0}\gamma_{i}-\gamma_{i}\gamma_{0}]=i\gamma_{0}\gamma_{i}=-i\alpha_{i}\nonumber\\&&
\sigma_{ij}=\frac{i}{2}[\gamma_{i}\gamma_{j}-\gamma_{j}\gamma_{i}]=i\gamma_{i}\gamma_{j}=\epsilon_{ijl}\Sigma_{l}
\eea
and our $\gamma_5$, $\hat{\beta}$ and $\gamma^{j}$ choice for convenience is
\bea
&\gamma_{5}=\left(\begin{array}{cc}
0 & -1 \\
-1 & 0\\ \end{array}\right)\;,\hat{\beta}=\gamma^{0}=\left(\begin{array}{cc}
1 & 0 \\
0 & -1\\ \end{array}\right)\;,\\&\quad\gamma^{j}=\left(\begin{array}{cc}
0 & \sigma^{j} \\
-\sigma^{j} & 0\\ \end{array}\right)\;,
\eea
where $0$ and $-1$ designate the corresponding $2\times2$ matrices \cite{silenko} and $\sigma^{j}$ are the Pauli matrices obeying the relation $\{\sigma^i\sigma^j+\sigma^j\sigma^i\}=-2g^{ij}$.
Hence, we may write the equation (\ref{eq:01}) as
\bea
\Bigl[i\gamma^\mu\partial_\mu+\mu(i\vec{\alpha}\cdot\vec{E}-\vec{\Sigma}\cdot\vec{B})-id(i\vec{\alpha}\cdot\vec{E}-\vec{\Sigma}\cdot\vec{B})\gamma_5\nonumber\\ -ig(\upsilon_{0}(\vec{\alpha}\cdot\vec{B})-\,\vec{\alpha}\cdot(\vec{\upsilon}\times\vec{E})+\vec{\upsilon}\cdot\vec{B})-m\Bigl]\psi=0\;,\label{EM}
\eea
where $\upsilon_{0}$ and $\vec{\upsilon}$ are time-like and space-like component of the coefficient $\upsilon_{\mu}$ respectively. We defined the matrices into Dirac standard representation:
\begin{eqnarray*}
&\vec{\alpha}=\hat{\beta}\vec{\gamma}=\left(\begin{array}{cc}
0 & \vec{\sigma} \\
\vec{\sigma} & 0\\ \end{array}\right)\;,\qquad
\vec{\Sigma}=\left(\begin{array}{cc}
\vec{\sigma} & 0 \\
0 & \vec{\sigma}\\ \end{array}\right)\;,\\ &\vec{\Pi}=\hat{\beta}\vec{\Sigma}=\left(\begin{array}{cc}
\vec{\sigma} & 0 \\
0 & -\vec{\sigma}\\ \end{array}\right)\;.
\end{eqnarray*}
Therefore, the equation (\ref{EM}) is write as
\bea
H\psi=[\vec{\alpha}\cdot\vec{\pi}+\mu\,\vec{\Pi}\cdot\vec{B}+id\,\vec{\Pi}\cdot\vec{E}
+g\,\hat{\beta}\vec{\upsilon}\cdot\vec{B}+\hat{\beta}m]\psi
\eea
where $\vec{\pi}=-i(\vec{\nabla}+\hat{\beta}(\mu\,\vec{E}+g\,(\vec{\upsilon}\times\vec{E})-d\,\vec{B}-g\,\upsilon_{0}\vec{B}))$. At this point, we want to find the nonrelativistic approach of our theory. Hence, we may use the Foldy-Wouthuysen (FW) transformation for Dirac spinor \cite{fw},  that has the goals of simplifies the transition from the semiclassical description  to the classical limit of relativistic quantum mechanics. In accordance with the work \cite{nosso}, we may obtain following Hamiltonian for the present case:
\bea\label{eq:03}
\hat{H}\approx \hat{\beta}\Bigl[m-\frac{1}{2m}\Bigl(\vec{\nabla}-i\vec{A}\Bigl)^{2}+ A_{0}\Bigl]
\eea
with
\bea
\vec{A}&=&\hat{\beta}\big(\mu(\vec{\Sigma}\times\vec{E})+g\vec{\Sigma}\times (\vec{\upsilon}\times\vec{E})-d(\vec{\Sigma}\times\vec{B})\nonumber\\&-&g\,\upsilon_{0}(\vec{\Sigma} \times\vec{B})\big)
\eea
and
\bea
A_{0}&=&-\frac{\mu^2}{2m}\vec{E}\,^2-\frac{d^2}{2m}\vec{B}\,^2+\frac{g^{2}}{2m}(\vec{\upsilon}\times\vec{E})^{2}+\frac{g^{2}\upsilon_{0}^{2}}{2m}\vec{B}^{2}\nonumber\\&-&\frac{\mu\hat{\beta}}{2m}\vec\nabla\cdot\vec{E}+\frac{d\hat{\beta}}{2m}\vec\nabla\cdot\vec{B}-\frac{ig\,\hat{\beta}}{2m}\vec{\nabla}\cdot(\vec{\upsilon}\times\vec{E})\nonumber\\&+&\frac{ig\,\hat{\beta}\upsilon_{0}}{2m}\vec{\nabla}\cdot\vec{B}+\mu\,\vec{\Pi}\cdot\vec{B}+id\,\vec{\Pi}\cdot\vec{E}
+g\,\hat{\beta}\vec{\upsilon}\cdot\vec{B}
\eea
The expression (\ref{eq:03}) is the nonrelativistic quantum hamiltonian for four-components fermions. However, for several applications in quantum mechanics, we may write (\ref{eq:03}) for two-components fermions in the form
\be\label{ham}
H= \frac{1}{2m}\Bigl(\vec{\nabla}-i\vec{a}\Bigl)^{2}+a_{0}\;,
\ee
that is similar to the interaction of a particle with the electric and magnetic fields minimally coupled to a non-Abelian gauge field with potential $a_\mu$, where
\bea\label{eq:04}
a_{0}&=&-\frac{\mu^2}{2m}\vec{E}\,^2-\frac{d^2}{2m}\vec{B}\,^2+\frac{g^{2}}{2m}(\vec{\upsilon}\times\vec{E})^{2}+\frac{g^{2}\upsilon_{0}^{2}}{2m}\vec{B}^{2}\nonumber\\&-&\frac{\mu}{2m}\vec\nabla\cdot\vec{E}+\frac{d}{2m}\vec\nabla\cdot\vec{B}-\frac{ig}{2m}\vec{\nabla}\cdot(\vec{\upsilon}\times\vec{E})\nonumber\\&+&\frac{ig\upsilon_{0}}{2m}\vec{\nabla}\cdot\vec{B}+\vec{\mu}\cdot\vec{B}+i\vec{d}\cdot\vec{E}
+g\,\vec{\upsilon}\cdot\vec{B},
\eea
and $\vec{a}=\vec{\mu}\times\vec{E}+\vec{g}\times (\vec{\upsilon}\times\vec{E})-\vec{d}\times\vec{B}-\upsilon_{0}(\vec{g}\times\vec{B})$, with $\vec\mu=\mu\vec\sigma$, $\vec{d}=d\vec\sigma$, $\vec g=g\vec\sigma$ and $\vec{\sigma}=(\sigma_1,\sigma_2,\sigma_3)$, $\sigma_i$ $(i=1,2,3)$ are the $2\times2$ Pauli matrices.    
The Hamiltonian (\ref{ham}) describes a system formed by a neutral particle that possesses permanent magnetic and electric dipole moments, and a constant parameter which control the Lorentz-symmetry violation in the presence of a electric and magnetic fields.

Now, we will study the nonrelativistic quantum dynamics of a particle corresponding to the Hamiltonian (\ref{ham}) which describes the physical situation, for example, the influence of Lorentz-symmetry violation parameters $\vec{\upsilon}_{0}\neq 0$ and $\vec{\upsilon}\neq 0$ on the dipole dynamics $\mu\neq 0$ and $d\neq 0$. Therefore, considering the expression (\ref{ham}), only by the Pauli-like term:
\bea
H=-\frac{1}{2m}\Bigl(\vec{\nabla}-i\vec{a}\Bigl)^{2}\label{eq:05}
\eea
where the other terms of (\ref{ham}) do not contribute to quantum phases due the choice of specific dipole-field configurations. The Schr\"odinger equation for this problem can be written as
\bea
-\frac{1}{2m}\Bigl(\vec{\nabla}-i\vec{a}\Bigl)^{2}\Psi=E\Psi\;,\label{eq:06}
\eea
To obtain the quantum phase we use the ansatz $\Psi=\Psi_{0}e^{\phi}$, where $\Psi_0$ is the free particle wave function without the influence of Lorentz-symmetry violation parameter, and $\phi$ is the phase that the particle wave function  accumulates due to the presence of the external fields and the Lorentz-symmetry violation parameter. The phase shift is calculated from the eq. (\ref{eq:06}) as a change in the classical action of the system, and is given by 
\bea
\phi&=&i\oint\,[\vec{\mu}\times\vec{E}+\vec{g}\times (\vec{\upsilon}\times\vec{E})\nonumber\\&-&\vec{d}\times\vec{B}-\upsilon_{0}(\vec{g}\times\vec{B})]\cdot d\vec \ell\;.\label{eq:07}
\eea
This phase represents the quantum phase acquired by a  neutral particle  the possess an electric and magnetic dipole moments in the presence of a parameter that controls a Lorentz-symmetry violation. Considering the expression (\ref{eq:07}) in the case where we have the absence of the parameters $\upsilon_{0}=0$ and $\vec\upsilon=0$ , this phase was studied by Anandan \cite{AN}.

Now, we analyze some special case. The first phase is the Aharonov-Casher quantum phase in the presence of Lorentz violation parameter. We consider in this case the absence of the magnetic field $\vec B=0$, we have 
\bea
\phi_{AC}=i\oint\,\Bigl[\vec{\mu}\times\vec{E}+\vec{g}\times (\vec{\upsilon}\times\vec{E})\Bigl]\cdot d\vec \ell\;.\label{eq:09}
\eea
The contribution $\vec{g}\times (\vec{\upsilon}\times\vec{E})$ is the factor that determines the  correction of Aharonov--Casher phase due to space-like component of Lorentz-symmetry violation parameter. However, the terms $\vec{\mu}\times\vec{E}$ and $\vec{g}\times (\vec{\upsilon}\times\vec{E})$ do not simultaneously contribute to the topological phase shift, since $\vec\mu$ and $\vec E$ must be orthogonal and $\vec g$ is by definition parallel to $\vec\mu$.  Using the BAC-CAB identity $\vec{g}\times (\vec{\upsilon}\times\vec{E})=(\vec g\cdot\vec E)\vec\upsilon-(\vec g\cdot\vec\upsilon)\vec E$, we see that the first term in second hand vanishes and the second term do not contribute to the phase in eq. (\ref{eq:09}), since $\vec\nabla\times\vec E=0$. Therefore, we obtain only the original Aharonov--Cahser phase $\phi_{AC}=i\oint\,(\vec{\mu}\times\vec{E})\cdot d\vec\ell$. However, if we consider a unusual electric field generated by a density of magnetic current $\vec\nabla\times\vec E=-\vec J^\mathrm{m}$, we have a non zero contribution from the term $\vec{g}\times (\vec{\upsilon}\times\vec{E})$ to the topological phase. In this case, the contribution from the term $\vec\mu\times\vec E$ to phase shift vanishes. Thus the topological phase is given by $\phi=-i(\vec g\cdot\vec\upsilon)\oint\vec E\cdot d\vec \ell=i(\vec g\cdot\vec\upsilon)\int_S\vec J^\mathrm{m}\cdot d\vec S$.

Now in case of absence of electric field $\vec E=0$, that correspond to the generalization of the He-McKellar-Wilkens phase in the presence of a parameter that break the Lorentz symmetry, we have 
\bea
\phi_{HMW}=-i\oint\,\Bigl[\vec{d}\times\vec{B}+\upsilon_{0}(\vec{g}\times\vec{B})\Bigl]\cdot d\vec \ell\;.\label{eq:08}
\eea
The presence of term $\upsilon_{0}(\vec{g} \times\vec{B})$, determines the correction to He--McKellar--Wilkens phase due the Lorentz-symmetry violation. We can write the expression (\ref{eq:08}) as
\bea
 \phi_{HMW}=-i\oint\,(\vec{\mathcal{D}}\times\vec{B})\cdot d\vec \ell,\label{eq:08.1}
\eea
where $\vec{\mathcal{D}}=[{d+g\upsilon_0}]\vec{\sigma}$ is the electric dipole moment modified by time-like component of Lorentz-violation parameter. Notice that the presence of the parameter of violation increase the electric moment of dipole of the particle introducing a new contribution in the He-McKellar-Wilkens phase. This contribution can be investigated using  interferometry of neutral particle that not possesses permanent electric dipole moments and observing the  increase of the magnitude electric polarizability of the particle. 

Now, we want to characterize the background of Lorentz-symmetry violation. In this sense we consider a neutral non polarized particle in the presence of a electric field $\vec E$ and the Lorentz-symmetry violation parameter $\vec\upsilon$. Thus, the phase in eq. (\ref{eq:09}) is reduced to $\phi=i\oint\vec{g}\times (\vec{\upsilon}\times\vec{E})\cdot d\vec \ell=i\oint(\vec g\cdot\vec E)\vec\upsilon\cdot d\vec\ell-i\oint(\vec g\cdot\vec\upsilon)\vec E\cdot d\vec\ell$. In order, the integral $\oint(\vec g\cdot\vec E)\vec\upsilon\cdot d\vec\ell$ will contribute to the phase only if the parameter $\vec\upsilon$ has a non null rotational. In this sense, we may admit that this parameter would be associated to a magnetic field generated by a density of electric current, $\vec\upsilon\equiv\vec B$ and $\vec\nabla\times\vec B=\vec J^\mathrm{e}$. Therefore, we obtain the phase in a interesting symmetric form $\phi=i(\vec g\cdot\vec E)\int_S\vec J^\mathrm{e}\cdot d\vec S+i(\vec g\cdot\vec B)\int_S\vec J^\mathrm{m}\cdot d\vec S$.

In this letter we investigated the a neutral quantum particle in the presence of an external magnetic and electric field and the Lorentz  violation parameter. We  demonstrated that a Lorentz-symmetry violation parameter contribute to change the geometric phase. The Aharonov-Casher effect in this model was investigated and we demonstrate that Lorentz-symmetry violation parameter not  contribute to Aharonov-Casher effect, but considering the presence of currents of a magnetic charge we found new contribution to Aharonov-Casher effect due to Lorentz-symmetry violation parameter. The  He-McKellar-Wilkens effect also has been investigated in this context and have demonstrated that Lorentz-symmetry violation contribute to increase the electric dipole of the particle.

{\bf Acknowledgments.} This work was partially supported by Funda\c c\~ao de Amparo \`a Pesquisa do Estado de S\~ao
Paulo (FAPESP), Conselho Nacional de Desenvolvimento Cient\'{\i}fico e Tecnol\'{o}gico (CNPq) CAPES/PROCAD.

\end{document}